# Experimental assessment of the effective friction at the base of granular chute flows on smooth incline

Olivier Roche[1]*, Siet van den Wildenberg[1], Alexandre Valance[2], Renaud Delannay[2], Anne Mangeney[3], Lucas Corna[1], Thierry Latchimy[4]

[1] Université Clermont Auvergne, CNRS, IRD, OPGC, Laboratoire Magmas et Volcans, F-63000 Clermont-Ferrand, France.
[2] Institut de Physique de Rennes, UMR CNRS 6251, Université de Rennes 1, Campus de Beaulieu Bâtiment 11A, 263 av. General Leclerc, 35042 Rennes CEDEX, France.
[3] Seismology Group, Institut de Physique du Globe de Paris, Université Paris Diderot, Sorbonne Paris Cité, 1 rue Jussieu, 75005 Paris, France.
[4] Université Clermont Auvergne, CNRS, UMS 833, OPGC, Aubière, France,

* Corresponding author (olivier.roche@uca.fr)

We report on direct measurements of the basal force components for granular material flowing down a smooth incline. We investigate granular flows for a large range of inclination angles from θ=13.4° to 83.6° and various gate opening of the chute. We find that the effective basal friction coefficient, $\mu_B$, obtained from the ratio of the longitudinal force to the normal one exhibits a systematic increase with increasing slope angle and a significant weakening with increasing particle hold-up H (the depth-integrated particle volume fraction). At low angles, the basal friction is slightly less than or equal to tanθ. The deviation from tanθ can be interpreted as a contribution from the side-wall to the overall friction. At larger angles, the basal friction $\mu_B$ saturates at an asymptotic value that is dependent on the gate opening of the chute. Importantly, our data confirm the outcomes of recent discrete numerical simulations. First, for steady and fully developed flows as well as for moderately accelerated ones, the variation of the basal friction can be captured through a unique dimensionless number, the Froude number Fr, defined as Fr=$\overline{U}/(gH\cos\theta)^{1/2}$, where $\overline{U}$ is the mean flow velocity. Second, the mean velocity scales with the particle hold-up H with a power exponent close to 1/4, contrasting with the Bagnold scaling ($\overline{U} \sim H^{3/2}$).

**I. INTRODUCTION**

Gravity-driven granular flows occur commonly in many industrial and geophysical contexts [1,2]. These flows are controlled by collisions and enduring frictional contacts between the solid particles and between the particles and the flow boundaries. Limiting types of granular flows are slow dense flows on substrate with slope angle close to the repose angle of the granular material, which are dominated by frictional contacts, and rapid dilute flows generally on steep slopes and which are dominated by collisions. Over the past decades, intensive research has been conducted to formulate robust constitutive laws for granular flows. Slow and dense granular flows on bumpy substrates (i.e. made with glued particles) have been shown to be well described by the frictional µ(I) rheology, which has met with considerable success [3-6]. This rheological law tells us that the effective friction coefficient, defined as the ratio of the shear stress over the normal stress, is a unique function of the inertial number I, defined as $I = \dot{\gamma}d/(P/\rho)^{1/2}$, where d is the particle diameter, $\dot{\gamma}$ the shear rate, P the pressure and ρ the particle density [1]. However, the role of the boundary conditions and their effects on the rheology is much less documented. In particular, flows on flat boundaries where slip velocities can be significant have received much less attention [7,8].

Experimental works involving smooth boundaries are indeed very few [9-11]. Experiments by Louge et al. [9] were conducted at low slope angles between 15° and 20°. These authors observed dense flows that became steady and fully-developed (SFD) before leaving the chute. The flows consisted of a basal mono-layer of rolling particles (which may slide or not) on the flat substrate and topped with the rest of the granular material. Though the frictional sidewalls affected the flow velocity, their relative contribution in the force balance was supposed to be negligible. Heyman et al. [11] conducted similar experiments with a smaller chute width W (i.e., W=40d against W=68d in [9]). They investigated larger inclination angles and showed that, between 20° and 30°, it is possible to observe flows that still become SFD before leaving the chute but develop secondary flows (rolls) that manifest themselves as longitudinal vortices. The longitudinal rolls were also observed in chute flow experiments with bumpy boundaries [12-14]. Holyoake and McElwaine [10] investigated even steeper angles between 30° and 50° in a 4 m long and 0.25 m wide chute. The granular flows were still accelerating at the exit of the chute and revealed striking properties. The authors observed a depletion of particles at the side-walls, which was probably the signature of the existence of a supported flow regime as evidenced in discrete element simulations and discussed below [14,15].

Discrete element simulations by Brodu et al. [15,16] revealed further details on the properties of flows on a smooth incline. The advantage of simulations is that they allow investigating SFD flows at arbitrary high angles that are not accessible to experiments unless equipped with unpractically long channels. These simulations, achieved with a finite chute width W=68d, reproduced longitudinal rolls at inclination angles of 20-50° [14,16]. They also revealed supported flows at angles typically larger than 30° [14,15,17]. A supported flow is characterized by a dense core bordered by agitated and strongly sheared dilute regions at boundaries. The emergence of the dense core is caused essentially by inelastic collapse of the expanded granular flow and hence is favored by low elastic restitution coefficient of the particles. These supported flows exhibit properties that are significantly different from homogeneous dense flows. For example, the mean velocity is found to scale as $H^\alpha$, with an exponent $\alpha$ of the order of ¼, where H is the particle hold-up defined as the depth-integrated particle volume fraction $\phi$,

$$H = \int_0^\infty \phi(z)dz \qquad (1)$$

H is proportional to the mass M of the flowing material per unit basal area (M=$\rho \times$H, where $\rho$ is the density of the particles). Note that H has the dimension of a length and may be expressed as well as H=$<\phi>\times$h, where $<\phi>$ is the depth-averaged particle volume fraction and h the flow height. This scaling contrasts markedly with the scaling law for dense flows over bumpy base where the mean velocity scales as $H^{3/2}$ [18].

One fundamental issue concerning flows on smooth inclines is whether they can be described by the µ(I) rheology. Numerical simulations [15] and experiments [19] indicate that dense flows on flat base, that is with moderate inclinations (typically below 20°), are reasonably well captured by the µ(I) rheology. However, this rheology does not provide any predictions of the slip velocity at the base nor indicates how the latter varies with the flow parameters (H and θ). At higher inclinations, the flows exhibit large heterogeneities in particle density as discussed above and they are by nature outside the framework of the µ(I) rheology. Granular flow experiments on steep slope by Holyoake and McElwaine [10] confirm that the µ(I) rheology is not able to capture correctly the dynamics of these flows.

Another important issue concerns the characterization of the boundary conditions at the flow base. Recent discrete numerical simulations [20] indicate that the basal slip velocity is simply related to the effective basal friction $\mu_B$. Zhu et al. [20] indeed showed that the basal friction is a monotonic function that depends on a unique dimensionless number, Fr, defined from an analog of a Froude number as $Fr=U_B/(gH\cos\theta)^{1/2}$, where $U_B$ is the basal flow velocity, H the particle hold-up and $\theta$ the inclination angle. For large Froude number, the basal friction eventually saturates at an asymptotic value corresponding to the microscopic friction coefficient ($\mu_m$) between the particles and the basal boundary. Interestingly, Zhu et al. [20] demonstrated in addition that the basal friction law $\mu_B(Fr)$ holds at a local scale and also for unsteady flows. These numerical outcomes have been however not checked experimentally.

An experimental confirmation of the aforementioned results requires direct measurements of the basal friction, which is itself a technological challenge. Very few attempts have been made in the context of rapid granular flows whereas these measurements are more common in slowly sheared granular systems [21-24] because less challenging. Indirect measurements have been achieved by Holyoake and McElwaine [10] for granular flows on steep slopes and by Faug et al. [19] in shallow granular flows with granular jumps. We are aware of only one attempt of direct measurement of the shear stress by Hungr and Morgenstern [25], but these authors estimated the normal stress from the flow thickness and density, thus producing potentially large uncertainties.

In the present study, we responded the challenge of direct measurement of the basal friction. We implemented a three-component force sensor within the basal substrate of a chute flow experiment, allowing simultaneous assessments of the normal and shear stresses at the flow base. We investigated granular flows on a smooth incline, within a large range of inclination angles from $\theta=13.4°$ to $83.6°$ and various depths. We reported on direct experimental measurements of the effective basal friction coefficient for flows reaching a SFD regime before leaving the chute, but also for flows accelerating along the whole chute. These measurements provide new insights about the effective basal friction of granular flows down a smooth plane and confirm the outcomes of the recent simulations by Zhu et al. [20].

The paper is organized as follows. In section 2 we describe the experimental set-up for the chute flow and how the force sensor is implemented. We present in section 3 the force measurements at the flow base as well as the properties of the granular flows. In section 4, we discuss our results and compare them to the numerical outcomes of Brodu et al. [15] and Zhu et al. [20]. We provide some concluding remarks in section 5.

## II. EXPERIMENTAL METHOD

The experimental device consisted of a 180 cm-long and 30 cm-wide inclined channel whose upper part was a reservoir from which particles were released to generate granular flows (Fig. 1). The particles were glass beads of diameter $d=1.5\pm0.1$ mm, density $\rho=2500$ kg/m$^3$, and repose angle ~22°. The channel and the reservoir were separated by a gap of ~1 mm, smaller than the particle size, in order to avoid mechanical coupling between the two parts of the device. The reservoir had a double gate system, with the inner gate in fixed position and setting a constant opening $h_g=2$, 4 or 6 cm while the outer gate was removed to release the particles. The channel had smooth boundaries, which were an aluminum plate at the base and Perspex plates at the lateral sides. The inclination could be varied from a horizontal to nearly a vertical position. Practically, we investigated slope angles between 13.4° and 83.6°. Our aim was to investigate a range of phenomena, from slow flows reaching a SFD regime before leaving the chute at low slope angles to rapid flows accelerating along the whole chute at steeper angles. The experiments were filmed with a high-speed video camera at 1000-5000 frame/s, either from above to measure the velocity and the acceleration of the flow front, or through the transparent sidewalls to investigate the height of the flows. The front velocity and the front acceleration at the location where the force sensor was implemented (see below) were determined from the data of the front position as a function of time along the channel. In all cases the front moved at constant speed without significant deformation, so that the front velocity was equal to the mean flow velocity [26]. When the front accelerated, the equality still hold since the front shape remained unchanged. Finally, the flow rate was measured via a weight scale at the channel outlet.

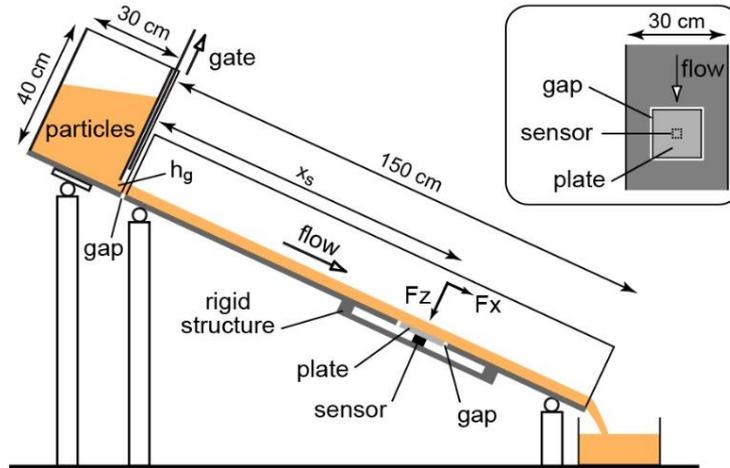

**FIG 1.** Sketch of the experimental device with the force sensor at distance $x_s$=98 cm from the reservoir. $F_x$ and $F_z$ are respectively the shear and the normal force components measured by the sensor (the transverse component $F_y$ is perpendicular to the image). The plate on top of the force sensor is shown in light grey. In the reservoir, $h_g$ is the opening of the inner gate. The gaps are not to scale. The inset is a view from above of the channel at the measuring plate.

We measured the forces at base of the granular flows with a 3-component piezoelectric sensor (Kistler™ 9317B), which permitted us to determine simultaneously the tangential shear ($F_x$) and transverse ($F_y$) force components as well as the normal force component ($F_z$). From these data, we could compute the ratio $F_x$ to $F_z$, which can be interpreted as an effective basal friction coefficient:

$$\mu_B = \frac{F_x}{F_z} \qquad (2)$$

The force measured by the sensitive element was proportional to the relative displacement between the upper and the lower parts of the sensor. The sensor had a response threshold <0.01 N, a rigidity of 190 N/µm ($F_x$, $F_y$) or 900 N/µm ($F_z$), and a proper frequency of 5 kHz ($F_x$, $F_y$) or 21 kHz ($F_z$). Considering the typical forces applied in our experiments, the high rigidity caused relative displacement of ~$10^{-2}$-$10^{-3}$ µm between the two parts of the sensor. We used a sampling frequency of 3 kHz for doing the force measurements. The system for measuring the forces consisted of a 15x15 cm flush-fitting aluminum plate inserted into the base of the channel and screwed on the upper part of the sensor whose lower part was connected to a rigid structure fixed to the channel base. The sensor was at the center of the plate and at a distance $x_s$=54 cm or 98 cm from the reservoir gate. A mean gap of ~150 µm between the margins of the plate and that of the insert in the channel base allowed the plate and the upper part of the sensor to move freely while preventing the particles from falling into the gap and blocking the plate. The gap was set by inserting 100 µm-calibrated steel sheets between the plate and the channel base before the system, including the plate, the sensor and the lower rigid structure, was set up. Then the steel sheets were removed.

We made tests to ensure for the reliability of our force measurements. We checked for accuracy in static configuration by placing an object of know weight P=2.64 N at the center of the plate while the channel was inclined at θ=20-30°. The measured forces $F_x$ and $F_z$ were respectively equal to the tangential and normal weight components $P_x=P\times\sin\theta$ and $P_z=P\times\cos\theta$ within an error of 1-2 %. We also investigated the drift of the signal delivered by the sensor and we found that it was detectable only after several tens of seconds. As the flows in experiments lasted less than 5 s, except at low slope angles of 13.4° and 15.2° where they could last up to 15 s, we considered that signal drifting was negligible for the force measurements. Notice that the plate was wider than the sensor and that a bending torque might occur in case of significant variation of thickness of the granular material. Observations showed, however, that variations of flow thickness across the plate were negligible or were at least within the range of error of thickness estimates.

## III. RESULTS

We chose to investigate the influence of both the gate opening and distance travelled down the channel prior reaching the sensor on the flow regime and associated basal forces. Therefore, we conducted a first series of measurements at given distance $x_s$=54 cm while $h_g$ was varied from 2 cm to 6 cm, and in a second series, the gate opening was kept constant ($h_g$=2 cm) while the sensor was set up at further downstream distance $x_s$=98 cm. We present first the basal force measurements and then the main flow properties (front velocity and acceleration, particle hold-up, and mass flow rate) that are useful for the interpretation of the force data. Note that our analysis focuses on flow properties in the main flow direction and disregards secondary flows (i.e. longitudinal vortices) that might develop within the main flow. According to the numerical simulations by Brodu et al. [15], the magnitude of the secondary flows are small in comparison with the main flow velocity so they can be neglected when dealing with averaged quantities.

### A. Measurement of basal forces

Examples of signals delivered by the force sensor are presented in Fig. 2. In all experiments, the values of $F_x$ and $F_z$ increased when the flow front arrived at the plate, then they were approximately constant for a while (plateau) as the flow thickness was constant, and they finally decreased as the flow thickness decreased and the particles eventually evacuated the channel. In contrast, the time-averaged mean value of $F_y$ remained close to zero for the whole flow duration. At low slope angle (i.e., at θ=13.4°), the force plateau of $F_x$ or $F_z$ was surprisingly of relatively short duration (less than 1 s) compared to the flows at higher angles. After the plateau, the flow thinned very slowly probably because we were very close to the jamming transition. In contrast, the plateau was much better defined at steeper angles and the flow thickness decreased rapidly after the maximum forces were reached. The time-averaged values of $F_x$ and $F_z$ in the plateau decreased notably with the channel slope angle (Fig. 3). Importantly, the effective basal friction coefficient $\mu_B=F_x/F_z$ was constant during the force plateau but also at earlier or later stages provided the forces were not too low. Notice that the relative fluctuations of the values of the three force components around their mean values was ~0.05-0.10 N and did not reveal any significant trend at increasing slope angles. In contrast, similar fluctuations for $F_x/F_z$ increased by one order of magnitude because both $F_x$ and $F_z$ decreased significantly while the force fluctuations were still around the same amplitude (Fig. 2).

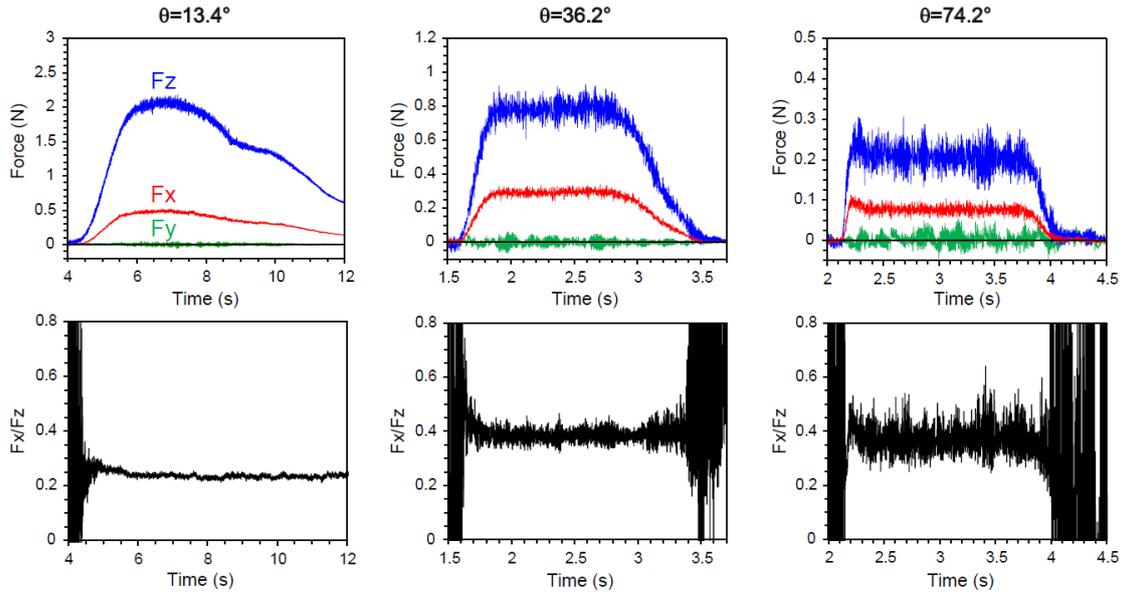

**FIG 2.** Examples of measurement with the force sensor at $x_s$=98 cm and $h_g$=2 cm, for flows at slope angle θ=13.4°, θ=36.2° and θ =74.2°, respectively. The graphs show the force components $F_x$ (red), $F_y$ (green) and $F_z$ (blue) as well as the basal effective friction coefficient $\mu_B=F_x/F_z$ as a function of time. Note that although the flow at θ=13.4° lasted more than 20 s, the stationary regime is less than 1s.

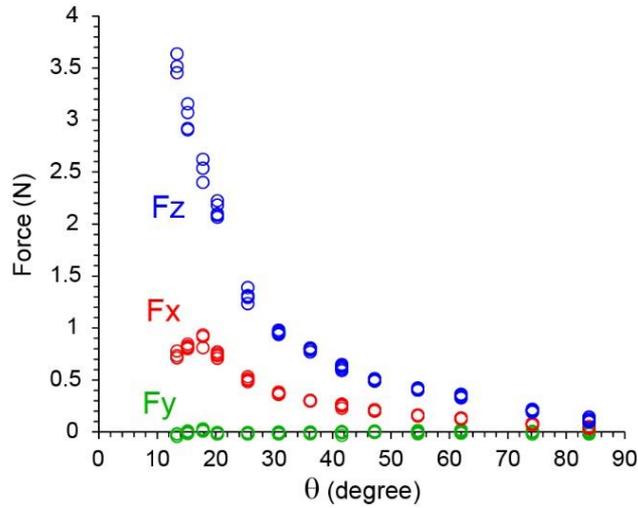

**FIG 3.** Mean values of the force components $F_x$, $F_y$ and $F_z$ in the force plateau (see Fig. 2) as a function of the slope angle θ for measurements at $x_s$=98 cm and $h_g$=2 cm.

The effective basal friction coefficient $\mu_B=F_x/F_z$ during the force plateau is shown in Fig. 4 as a function of the slope angle. We present four different data sets: three sets obtained at a downstream position $x_s$=54 cm for three different gate openings ($h_g$=2, 4 and 6 cm) and an additional set at a further downstream position $x_s$=98 cm for a single gate opening ($h_g$=2 cm).

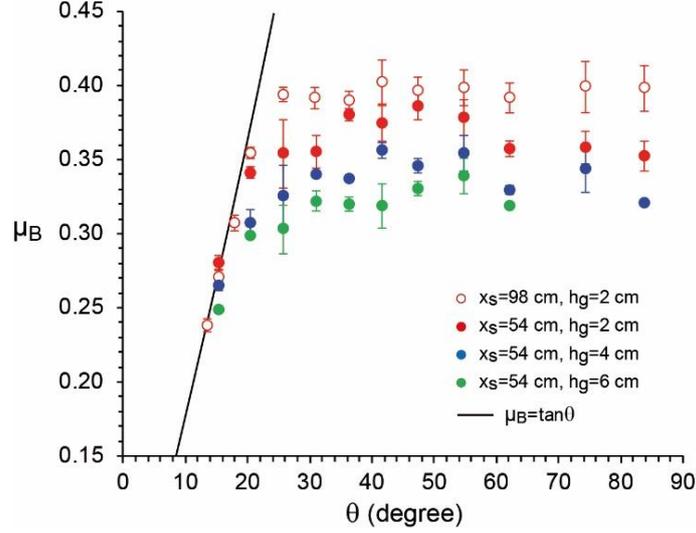

**FIG 4.** Effective basal friction coefficient $\mu_B = F_x/F_z$ as a function of the channel slope angle $\theta$. Four data sets are shown: three sets at a downstream distance $x_s = 54$ cm for gate openings $h_g = 2, 4$, and $6$ cm, respectively, and one set at $x_s = 98$ cm for a single gate opening $h_g = 2$ cm.

At $x_s = 54$ cm and for a fixed gate opening, the basal friction $\mu_B$ increases monotonically with increasing angle and eventually saturates at high angle. The asymptotic value of the basal friction, $\mu_{B,max}$, clearly decreases with increasing gate opening: $\mu_{B,max} \approx 0.38$ for $h_g = 2$ cm whereas $\mu_{B,max} \approx 0.32$ for $h_g = 6$ cm. At low angle and small $h_g$, the basal friction is found to match the tangent of the inclination angle but it deviates rapidly from this trend for increasing angle. For $h_g = 2$ cm, the deviation occurs for $\theta \geq 20.3°$ while for larger gate opening ($h_g = 4$ and $6$ cm), the deviation is observed sooner for $\theta \geq 15.2°$. The magnitude of the deviation thus increases with increasing gate opening. We can anticipate here that this result can be interpreted as a growing contribution of the side-wall friction when the flows get thicker.

Finally, the data sets obtained at two different downstream distances but same gate opening $h_g = 2$ cm reveal other interesting outcomes. At low angle, both data sets provide similar friction coefficient values. However, at larger angles ($\theta \geq 20.3°$), we observe a slight but measurable difference: the friction coefficient values obtained at the further downstream distance (i.e., at $x_s = 98$ cm) are systematically greater than the ones found at $x_s = 54$ cm. As discussed in more details later on, the data differ as soon as the flow is not fully-developed at the smaller downstream distance. According to the inclination angle, the flow can be still accelerating at $x_s = 54$ cm while it is fully-developed at $x_s = 98$ cm, or the flow is accelerating at both downstream distances.

**B. Flow characteristics**

To understand the variation of the effective basal friction with the gate opening and inclination angle, it is necessary to characterize the features of the granular flows in terms of velocity, acceleration, particle (mass) hold-up and mass flow rate.

Flows of a coherent mass of particles were possible at slope angles $\theta > \sim 13°$, though some particles could roll individually on the channel base at lower slope angles. For comparison, tests showed that a solid block resting on a layer of beads organized in 2D compact ordered packing and in contact with the channel base could slide at $\theta > \sim 7°$ if the beads were free to roll or at $\theta > \sim 17.5°$ if the beads were glued to the block. This suggested that the microscopic friction angle between the glass particles and the smooth aluminum base was close to $17.5°$.

The flow front velocity $U_F$ and acceleration $a$ were measured from videos and are presented in Fig. 5 as a function of $\sin\theta$, which is the driving component of gravity. The measurements were made at two different downstream distances ($x_s$=54 and 98 cm). Let us first analyze the data obtained for the same gate opening $h_g$=2 cm but different downstream distances. At low angles, both measurements coincide indicating that the flow becomes SFD before the smaller downstream distance ($x_s$=54 cm). The front acceleration at both distances is insignificant (see Fig. 5b). At intermediate angles ($\theta\approx 20.3°$ or $\sin\theta\approx 0.35$), the front acceleration at $x_s$=54cm becomes measurable ($a/g\geq 0.05$) while at $x_s$=98 cm, it is still too small to be measured. In other words, the flow is still accelerating at $x_s$=54 cm but SFD at $x_s$=98 cm. At higher angles (i.e., $\theta\geq 25.7°$ or equivalently $\sin\theta\geq 0.43$), the flow is accelerating at both downstream distances. For these high angles, the front velocity measured at $x_s$=98 cm becomes significantly greater than the one evaluated at $x_s$=54 cm, indicating that the flow front acceleration cannot be disregarded between $x_s$=54 and 98 cm. Finally, it is important to note that for significant large angles $\theta\geq 40°$ (i.e, $\sin\theta\geq 0.65$), the front accelerations become comparable at both downstream positions. This indicates that for slope angles above $40°$, the flow do undergoes a quasi-uniform acceleration along the chute. Additionally, data obtained at $x_s$=54 cm for different gate openings indicate that the front velocity increases almost linearly with $\sin\theta$ and presents only a slight increase with increasing gate opening. Importantly, the acceleration becomes significant ($a/g\geq 0.05$) at $x_s$=54 cm and angles $\theta\geq 20.3°$ for all gate openings. Given the measurement accuracy, it is hard to tell whether the distance $x_{SFD}$ at which the flow becomes fully developed depends or not on the gate opening.

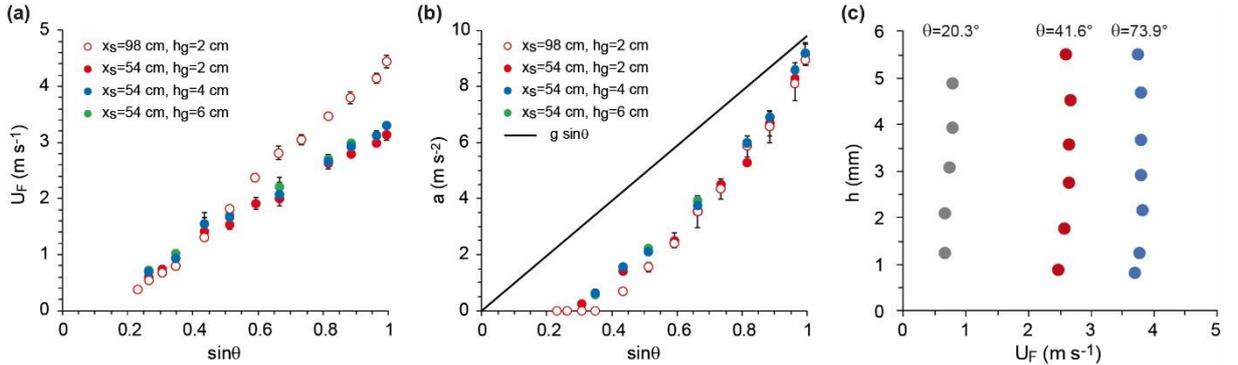

**FIG 5.** Flow front velocity $U_F$ (a) and acceleration $a$ (b) as a function of $\sin\theta$, measured at downstream distances $x_s$=54 and 98 cm and for gate openings $h_g$=2, 4 or 6 cm. (c) Velocity profiles measured at the side-walls for various inclination angles ($\theta$), at $x_s$=98 cm and for $h_g$=2 cm.

Several important issues concerning the front velocity $U_F$ should be addressed. The first one concerns its relationship with the mean flow velocity. As already mentioned above, in case of stationary front, there is a strict equivalence between the front velocity and the mean flow velocity [26]. In contrast, if the flow front is accelerating, the equality is not necessarily ensured. However, we checked through mass flow rate measurements (as detailed below) that the front velocity was a good approximation of the mean flow velocity even for accelerating flows. The second issue concerns the type of velocity profile obtained for such flows. As shown in Fig. 5c, the profiles look like plug-flow with an almost uniform velocity through the depth and a presumably large slip velocity at the base. We have not a high enough resolution to provide an accurate value of the slip velocity but we can reasonably conclude the mean flow velocity $\bar{U}$ (i.e., the depth-averaged flow velocity) and the slip velocity $U_B$ at the base are expected to be roughly equivalent. This is ascertained by discrete numerical simulations [15,20]. The latter also indicate that for such rapid flows on a smooth base, the shear is essentially localized at the base within a layer of a few particle diameter, which is compatible with the measured velocity profiles despite under-resolved.

An important relevant quantity that characterizes the flow is the particle hold-up H (or mass hold-up). This quantity is difficult to assess without force sensor and is therefore rarely measured. Force measurement provides useful pieces of information. First, it is a measure of the mass of the flow per unit area and is thus often used as a control parameter in numerical simulations. Second, when combined with the mean flow velocity $\overline{U}$, it gives us an estimation of the mass flow rate Q:

$$Q \approx \rho W H \overline{U} \qquad (3)$$

where W is the flow width and ρ the density of the particles. The particle hold-up is readily deduced from the normal force component $F_z$:

$$H = \frac{F_z}{\rho g\, A \cos\theta} \qquad (4)$$

where A is the area of the plate on the sensor. Fig. 6a presents the variation of the particle hold-up with the inclination angle for various gate openings ($h_g$=2, 4 and 6 cm) and at two different downstream positions ($x_s$=54 and 98 cm). The normalized particle hold-up H/d decreases with increasing slope angle, from ~4-11 at low inclinations to ~1-7.5 at the highest slope angles. The decreasing trend is less significant above 50 degrees and the particle hold-up seems to reach a plateau. The slight increase of the particle hold-up above 70 degrees is certainly due to measurements uncertainties. This is indeed confirmed by an alternative method for assessing the particle hold-up based on mass flow rate measurements, which show no increase of the particle hold-up at high angle (see square symbols in Fig. 6a). Interestingly, increasing the gate opening generates thicker flows and leads to an augmentation of the particle hold-up, as expected.

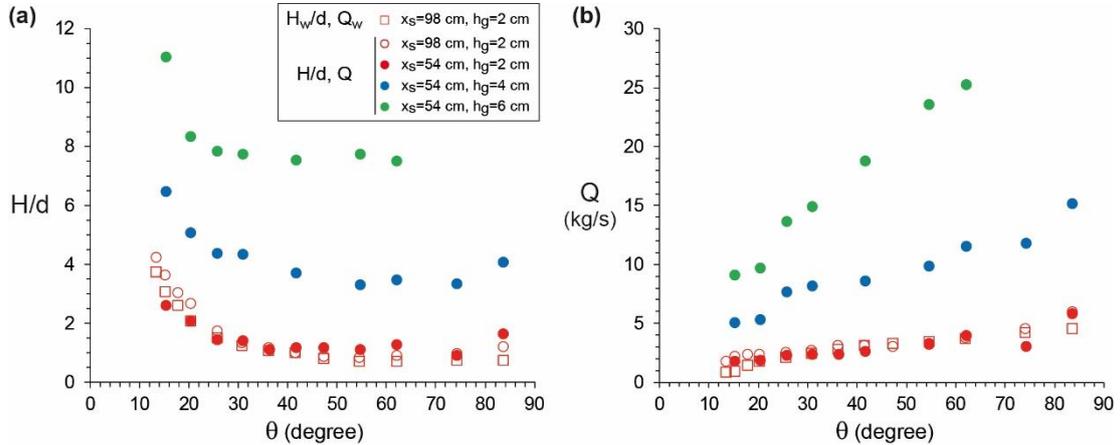

**FIG 6.** (a) Particle hold-up H (measured in grain diameter unit) as a function of the slope angle, given by Eq. 4, for various gate openings ($h_g$=2, 4 and 6 cm) and two different downstream positions ($x_s$=54 and 98 cm). Square symbols correspond to data inferred from Eq. 3 ($H_w$=$Q_w$/ρW$\overline{U}$, where $Q_w$ is the mass flow rate measured directly via a weight scale at the channel outlet). (b) Corresponding mass flow rates obtained from Eq. 3. Note that at $x_s$=98 cm and $h_g$=2 cm, the agreement between the flow rate Q deduced from Eq. 3 (circle symbols) and $Q_w$ measured directly via the weight scale (square symbols) is fairly good.

The determination of the particle hold-up allows us to have a quite straightforward estimation of the mass flow rate using Eq. 3 and assuming that the flow front velocity is a good approximation of the mean flow velocity. Fig. 6b presents the variation of the mass flow rate with the inclination angle for different gate openings. As expected, the flow rate increases both with increasing angle and gate opening. For $h_g$=2 cm, we checked that Eq. 3 provides similar results as those obtained from a direct measurement of the mass flow rate, $Q_w$, via a weight scale. This gives us confidence in our measurements for the particle hold-up H but also indicates that the flow front velocity is a good proxy for the mean flow velocity.

The decrease of the particle hold-up with the inclination angle can be due to a decrease of the flow height or a decrease of the particle concentration within the flow. We will see that this is caused mainly by a drastic diminution of the packing fraction. As stated in introduction, the depth-averaged packing fraction <$\phi$> is simply related to the particle hold-up via the flow height h:

$$<\phi> = \frac{H}{h} \qquad (5)$$

The estimation of the mean packing fraction thus requires the determination of the flow height h. The latter is defined as the position of the free surface of the flow with respect to the base of the channel. The free surface of the flow is relatively well-defined at the lowest slope angles for dense flows but much poorly-defined at large slope angles for more dilute flows (Fig. 7). For these dilute flows, the definition of the flow height requires the use of a clear and quantitative criterion. Several definitions may be suitable, either based on the critical height below which one finds a given percentage of the mass of the flow or defined as the position where the packing fraction falls below a critical value . We have employed a method based on the latter definition.

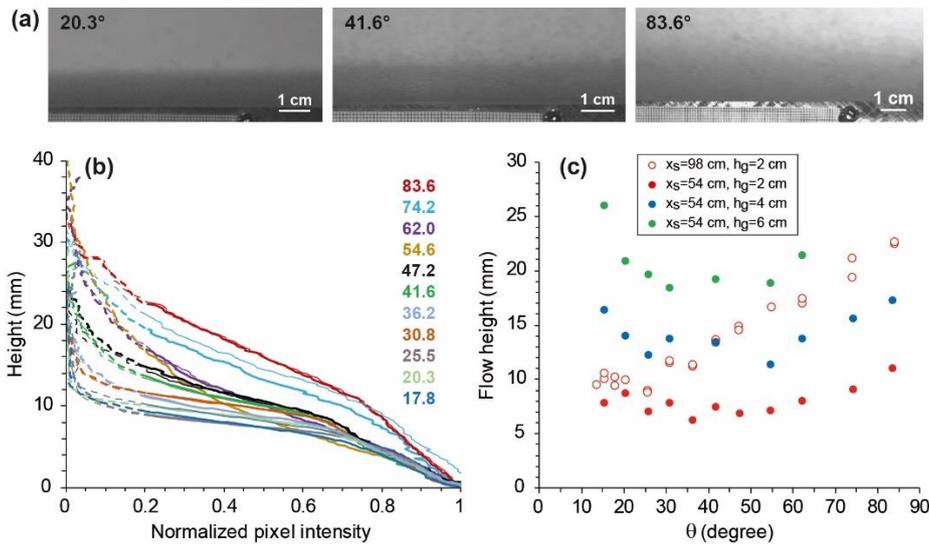

**FIG 7.** Flow structure and height. (a) Pictures of flows obtained by averaging of frames of size of ~8 cm from movies made through one of the transparent sidewall ($x_s$=98 cm, $h_g$=2 cm, flow from right to left). Numbers indicate the slope angles. (b) Normalized pixel intensity (after subtracting the background image) as a function of height above the channel base ($x_s$=98 cm, $h_g$=2 cm), where the value one corresponds to the maximum grey value. The flow height is defined as 20% of the maximum intensity as indicated by the transition from dashed to plain curves. Numbers on the right hand side of the graph indicate the slope angles. (c) Flow heights as a function of the slope angle, obtained from the procedure illustrated in (b).

Flow heights were obtained from the movies taken through the transparent sidewall. We considered the frames in which the flow was in the field of view, subtracted the background taken from a reference frame without any particles in the field of view, and summed the frames to obtain a vertical profile of gray level (i.e., perpendicular to the flow base). From these profiles, the height of the flow was defined as the vertical position where the gray level falls below 20% of the maximum grey value, which was observed at the flow base. The 20% threshold value was estimated from frames of SFD flows at low slope angles and whose free surface could be tracked accurately. However, we should be aware that the assessment of the flow height with this method is highly dependent on the value of the gray level threshold in particular for dilute flows. Consequently, data obtained for the latter should be taken with caution.

Fig. 7c shows that the flow height measured at $x_s=54$ cm decreases with increasing angles at low inclination ($\theta \leq 20°$), stabilizes at intermediate inclination angles ($20°<\theta<50°$) and finally increases at larger angles. Besides, the flow height increases with increasing gate opening. For intermediate angle, the flow height thus appears to be well correlated with the gate opening, so that $h \approx h_g/3$. At higher angle, we observe a clear expansion of the flow height but not greater than a factor 1.5. Measurements made at the downstream position $x_s=98$ cm with $h_g=2$ cm, however, reveals that the flows in the accelerating regime for $\theta \geq 25.7°$ dilate strongly while travelling downstream. In this regime, the flow height increases monotonically with the slope angle by a factor greater than 2.

The mean particle volume fraction can be now assessed using Eq. 5 (see Fig. 8). For a given gate opening, the mean packing fraction significantly decreases with increasing angle. At the lowest slope angles, it is close to the packing fraction $\phi_p \sim 0.645$, which was determined from the mass and the volume of a static bed of particles. At high angle (i.e., $\theta=83.6°$), the mean packing fraction goes down to very small values at $h_g=2$ cm: ~0.25 and ~0.05 at $x_s=54$ and 98 cm, respectively. These results confirm that the decrease of the particle hold-up with inclination angle is mainly due to a decrease of the packing fraction of the flow. In contrast, the flow height was shown to increase for increasing angle at large angle.

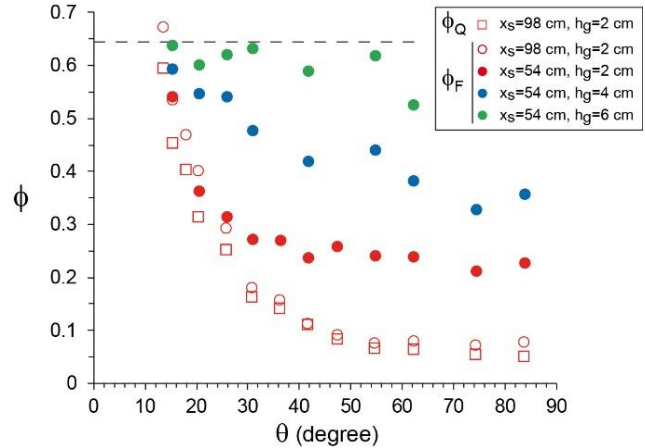

**FIG 8.** Mean particle volume fraction $<\phi>$ as a function of slope angle, gate openings ($h_g=2$, 4 and 6 cm) and two different downstream positions ($x_s=54$ and 98 cm). The dashed line indicates the packing fraction $\phi_p \sim 0.645$. At $x_s=98$ cm with $h_g=2$ cm, we also provide an alternative estimation of the packing fraction $<\phi_Q>$ based on the measurement of the mass flow rate $Q_w$ via a weight scale (see. Fig 6b): $<\phi_Q>=Q_W/\rho h W U$ (see Eq. 3). Both measurements indicate differences less than ~15%, except for angles larger than 60° where the volume fraction is smaller than 0.08.

## C. Summary

The basal forces and the effective basal friction coefficient we measured in our experiments depend both on the channel slope angle and on gate openings. The decrease of one order of magnitude of the components of the basal force with the slope angle (Fig. 3) is caused by a strong diminution of the mean particle volume fraction of the flow and thus a decrease of the particle hold-up (Fig. 6a). Flows at low slope angles have large particle concentrations approaching the close packing fraction $\phi_p \sim 0.645$ whereas the most rapid flows at the steepest angles have much smaller particle volume fractions down to $\phi < 0.10$ at $\theta > 40$-$70°$.

We also observed a general increase of the basal friction coefficient with increasing slope angle and decreasing gate opening. Importantly, the basal friction coefficient saturates at large slope angles. The saturation occurs as early as 20-25° for $x_s$=98 cm and $h_g$=2 cm, and around 40° in the other cases. The asymptotic value slightly increases with decreasing gate opening (0.32≤$\mu_{B,max}$≤0.38). According to Brodu et al. [15], this asymptotic value should be lower than the microscopic friction coefficient between the particles and the basal boundary, and should decrease with the particle hold-up. We observe this decrease but the corresponding effective basal friction angle 18.8≤$\phi_B$≤21.8° is slightly larger than the value of 17.5° required for sliding of a solid block with a glued basal layer of beads along the channel base. A better estimation of the microscopic friction coefficient would be necessary to conclude, here.

## IV. DISCUSSION

We present here an analysis of the force balance on our experimental flows in order to discuss the variation of the effective basal friction with the slope angle and gate opening. We discuss also about scaling laws for the flow front velocity and universal behavior of the effective basal friction.

### A. Force balance

We address the force balance on the flows according to the scheme shown in Fig. 9. We recall that that the sensor measures a deformation (thus a force) whatever the nature of the flow regime (i.e. steady or accelerated). We consider a flow slab of a finite length L and a height greater or equal to the flow height h, and spanning over the whole channel width W. We adopt a Lagrangian description, that is, we follow the system in course of its motion. Considering the driving gravitational force and both the basal and side-wall friction forces acting on a flow slab of mass m, we can write

$$m\gamma = mg\sin\theta - F_B - F_W \quad (6)$$

where $\gamma$ is the flow acceleration, and $F_B$ and $F_W$ are the basal and the side-wall friction forces, respectively. If the flow slab speed is constant, then the second Newton's law indicates that the resultant of the external forces (weight, basal friction, side-wall friction) should vanish. If not, the resultant gives information on its acceleration. Notice that this force balance disregards air drag and longitudinal pressure gradient forces. The latter can be safely neglected if the longitudinal gradient of the flow height is small, which is achieved for small enough L. In virtue of Eq. 2, the basal friction force is simply written as:

$$F_B = \mu_B mg\cos\theta \quad (7)$$

where $\mu_B$ is the effective basal friction coefficient and m=ρHWL is the mass of the flow slab. The side-wall friction force is expressed as [27,28]:

$$F_W = 2\mu_W L \int_0^\infty P(z)dz \quad (8)$$

where $\mu_W$ is the effective side-wall friction coefficient, P is the pressure within the flow (assumed to be isotropic, i.e., $P_{zz}=P_{yy}=P_{xx}=P$), and L is the length of the flow slab. Vertical momentum balance indicates that the basal pressure $P_B$ is equal to mgcosθ/(LW), with W the flow width, which provides us with a new expression of the side-wall friction force:

$$F_W = \mu_W mg\cos\theta \frac{Z}{W} \quad (9)$$

where

$$Z = (2/P_B) \int_0^\infty P(z)dz \quad (10)$$

The height Z can be interpreted as the height over which the flow experiences a significant friction from the side-walls. If the particle concentration is uniform throughout the flow depth, Z is easily estimated via Eq. 10 and is equal to the flow height h. If the flow is very dense, Z is thus close to H (h=H/<φ>), but always greater than H. In contrast, if the flow is dilute, Z is much greater than H. If the particle concentration is non-uniform throughout the flow as suspected for our rapid flows, the estimation of Z is possible only if the particle concentration profile is known, which is not the case here. As an example, if the particle concentration profile obeys a decreasing exponential law with a characteristic decay length l, Z is equal to twice this characteristic length l and is expected to be a significant fraction of the flow height h. As a consequence, Z is expected to much greater than H at large inclination angle.

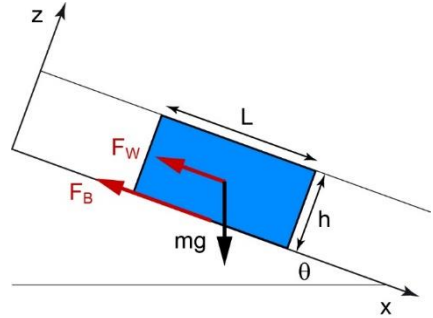

**FIG 9**. Force balance on the flows. The diagram shows a slab of a finite length L and height h, spanning over the whole channel width.

From Eqs. (6-9), the flow acceleration is given by
$$\frac{\gamma}{g\cos\theta} = tan\theta - \mu_B - \mu_W \frac{Z}{W} \quad (11)$$
As a first approximation, we can disregard in Eq. (11) the side-wall friction term, which is expected to be a small contribution with regard to the large channel width (W=200d), to estimate the flow acceleration $\gamma$. Taking advantage of this approximation, we get $\gamma \approx g \sin\theta$ (1-$\mu_B$/tan$\theta$) and we can compare it with direct measurements of the flow front acceleration $a$ given in Fig. 5b (see Fig. 10a). The agreement between $\gamma$ and $a$ is reasonably good, indicating that the basal friction force is indeed at first order the prevailing friction force. However, the deviation observed between $\gamma$ and $a$ can be attributed either to measurement uncertainties or to second order contribution from side-wall friction. In particular, when $\gamma$ is greater than $a$, the difference can be interpreted in terms of side-wall friction contribution, as discussed further below.

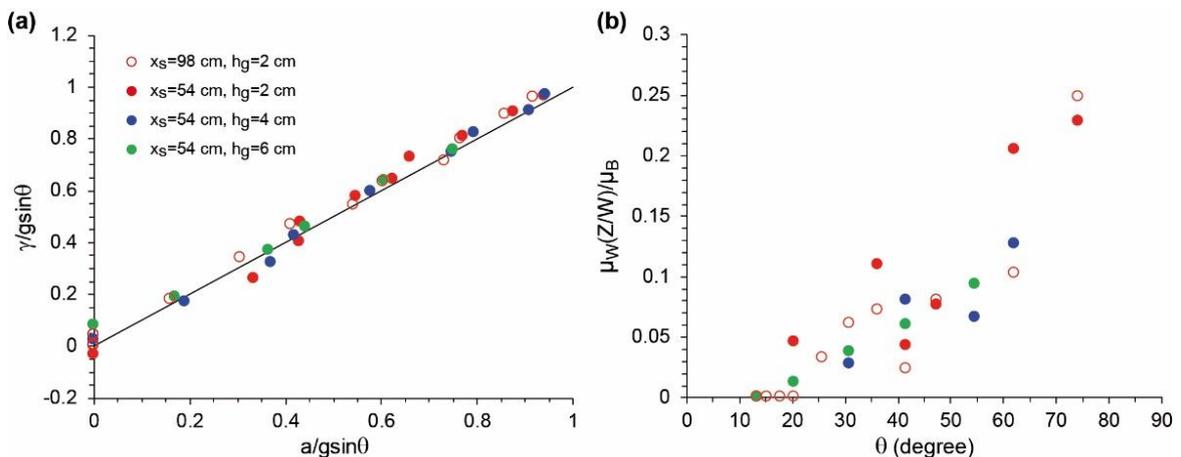

**FIG 10.** (a) Comparison between the flow acceleration $\gamma$ obtained from Eq. (11) (disregarding the side-wall friction term) and the measured flow front acceleration $a$. Both terms are normalized by g sin$\theta$. (b) Relative contribution of the side-wall friction $\mu_W$ (Z/W)/ $\mu_B$ as a function of the inclination angle. Data in (a) and (b) are shown for different gate openings $h_g$ and downstream distances $x_s$.

In a second step, it is tempting to try to get an estimation of the side-wall contribution term. At low inclination (e.g., at θ=17.8°), the flows are SFD at the location of the force measurements so that the acceleration is reduced to zero: the gravity term tanθ must therefore be balanced by the basal and side-wall frictions. For thin flows (i.e., $h_g$=2 cm), the gravity is balanced by the basal friction (tanθ≈$\mu_B$), meaning the side-wall friction is negligible. In contrast, for thicker flows (i.e., $h_g$=4 and 6 cm), the basal friction is significantly smaller than the tangent of the inclination, and the difference results from the contribution of side-wall friction (see Fig. 4). For arbitrary angles, it is possible to assess systematically the relative contribution of the side-wall friction with respect to the basal friction (i.e., $\mu_W(Z/W)/\mu_B$) as a function of the inclination angle, using Eq. (11) and setting γ=$a$ (see Fig. 10b). Although large fluctuations due to insufficient accuracy in the determination of $a$, the data indicate that the relative contribution of the side-walls roughly increases with increasing inclination angles and can rise up to ~25% of the basal friction at very large angle. One of the reasons of this increase observed at high inclination (i.e., θ>40°) can be attributed to the fact that the flows inflate at large angle (i.e., the flow height increases), resulting in a growing friction area at the side walls. The increase at smaller angle (i.e. 20°>θ>40°) is probably more tricky and may originate from the formation of a dilute and agitated granular gas layer at the walls as revealed by the numerical simulations [15,20], which induces an increase of the pressure at the side walls and consequently an augmentation of the side-wall friction.

**B. Scaling laws**

The flow conditions in our experiments are comparable with those used in the discrete simulations by Brodu et al. [15] involving flat and smooth base and side-walls. A first difference lies in the gap between the side-walls, which is much less in the simulations (W=68d) than in the experiments (W=200d). Second, the mechanical parameters characterizing the interactions between a grain and another grain and with the base or the lateral walls (such as the coefficients of restitution or the coefficients of friction) may be different. Third, in these simulations, SFD flows were obtained at arbitrary high angles of inclination. Experimentally, if SFD flows can be obtained, this could require very long chute. Despite these differences, it may be instructive to compare the experimental flows with those obtained in the discrete simulations.

The simulations of Brodu et al. [15] indicate that, for SFD flows, the mean flow velocity scales as $H^\alpha$, with α≈1/4. Our experimental results concerning SFD flows (i.e., θ≤20.3°) suggest a weak power law exponent, compatible with that of Brodu et al. [15] (Fig. 11). This contrasts with the Bagnold scaling, which predicts that $\bar{U}$~$H^{3/2}$.

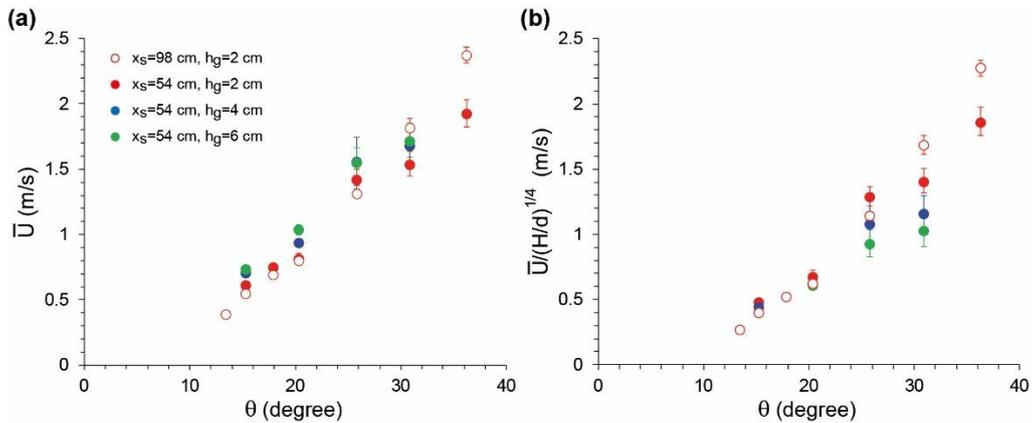

**FIG 11.** (a) Mean flow velocity $\bar{U}$ as a function of the inclination angle θ. (b) Mean flow velocity $\bar{U}$ rescaled by $(H/d)^{1/4}$ as a function of the inclination angle θ. The data collapse at θ≤20.3° for SFD flows.

Interestingly, recent simulations by Zhu et al. [20] show that for flows involving flat and smooth base and side-walls the effective basal friction is a unique function of a Froude number defined as

$$Fr = \frac{\bar{U}}{\sqrt{gH\cos\theta}} \quad (12)$$

In [20], the Froude number is defined from the slip velocity $U_B$ at the bottom, but as flows on smooth incline are essentially plug flows, the slip velocity $U_B$ is roughly equal to the mean flow speed $\bar{U}$. We checked whether this numerical finding holds as well for our flows. To test this, we used the front velocity instead of the basal velocity, which was not easily accessible. We re-plot our basal friction data given in Fig. 4 as a function of the Froude number (see Fig. 12a). Our data collapse reasonably well for SFD flows (i.e., Fr≤4 or θ≤20°). For moderately accelerating flows (i.e., 4<Fr<10), the collapse is less convincing but still roughly holds. At higher Froude number (i.e., Fr>10) corresponding to highly accelerating flows, we do not observe a collapse of the data. In numerical simulations, the friction law $\mu_B(Fr)$ was established for steady flows but also for unsteady flows. The reasons for which the collapse fails for strongly accelerating flows remains unclear. However, we may mention the possible influence of air drag which might be a source of additional friction.

The $\mu_B(Fr)$ curve has a shape similar to the $\mu(I)$ rheological curve for dense granular flows on rough substrates [1,29]: it first increases and then seems to saturate at large Froude number. However, we should make clear that these two laws have not the same status. As discussed in detail in [20], the $\mu_B(Fr)$ law provides a condition at the boundary relating the basal friction to the slip velocity, while the $\mu(I)$ law describes the internal rheology of the granular flow and informs about the strain rate within the flow. Importantly, the $\mu_B(Fr)$ curve provides a natural explanation for the decrease of the basal friction with the particle hold-up H. Indeed, in view of the velocity scaling law ($\bar{U} \sim H^{1/4}$), the Froude number decreases with increasing particle hold-up at a fixed inclination angle (Fr$\sim H^{-1/4}$). This results in a decrease of the basal friction since $\mu_B(Fr)$ is an increasing function of Fr.

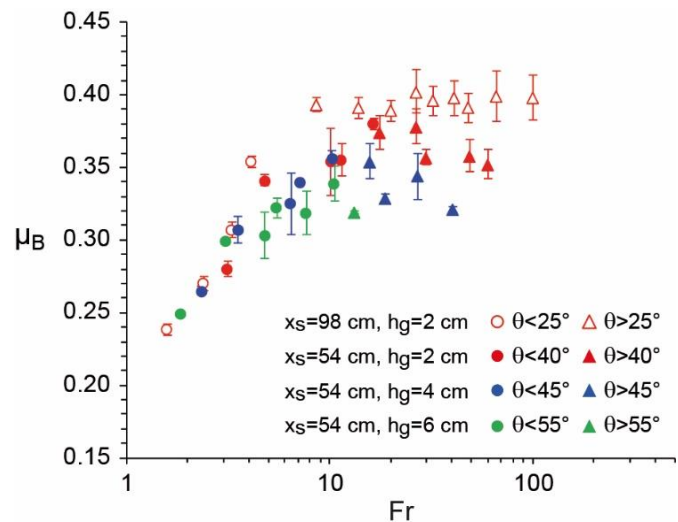

**FIG 12.** Effective basal friction coefficient (see Fig. 4) as a function of the Froude number Fr. The data collapse well for SFD flows (Fr≤4) and reasonably well for moderately accelerating flows (4<Fr<10) on a master curve.

## V. CONCLUSION

We have made measurements of the force components at the base of granular flows in an inclined channel with smooth boundaries. We investigated a large range of inclination angles from 13.4° to 83.6°. At low slope angle ($\theta \leq 20.3°$), after accelerating over a certain distance, the flows entered a SFD regime with packing fraction ranging from 0.4 to 0.6, whereas at larger angles the flows accelerated along the whole length of the channel and havd much lower packing fraction (down to less than 0.1). The measurements revealed a strong decrease of both the tangential shear and the normal force components at increasing slope angles mainly due to a diminution of the flow concentration and particle hold-up. The effective basal friction coefficient $\mu_B$ showed a systematic increase with the slope angle, in agreement with earlier findings from numerical simulations [15,20]. At low angles for thin flows, $\mu_B$ was close to the tangent of the slope angle. At higher angles corresponding to accelerating flows, $\mu_B$ increased at lower rate than $\tan\theta$, and it saturated at an asymptotic value of ~0.32-0.40 depending on the gate opening. Interestingly, the coefficient $\mu_B$ decreased significantly with increasing gate opening or particle hold-up.

A further analysis of the data allowed to understand some of these features. First, flows are essentially plug flows such that the mean flow velocity $\bar{U}$ is roughly equal to the basal slip velocity $U_B$. Second, we showed that for SFD flow regimes, the mean flow velocity $\bar{U}$ scales as $H^\alpha$ with $\alpha \approx 1/4$, in agreement with the discrete simulations by Brodu et al. [15]. This scaling contrasts markedly with the Bagnold scaling ($\bar{U} \sim H^{3/2}$). Third, we showed that for SFD and weakly accelerating flow regime, the effective basal friction coefficient can be reasonably captured by an increasing function of a dimensionless number, that is, the Froude number defined as $\bar{U}/(g H \cos\theta)^{1/2}$, in agreement with recent discrete simulations by Zhu et al. [20]. The $\mu_B(Fr)$ curve can be thought as a boundary condition for flows on smooth and flat boundary. Additionally, this curve gives a natural explanation for the decrease of the basal friction with the particle hold-up H. In virtue of the velocity scaling, the Froude number decreases with increasing particle hold-up at a fixed inclination angle, resulting in a decrease of the basal friction given that $\mu_B$ is an increasing function of Fr.

Our study demonstrates that the forces generated by granular flows at the boundaries can be assessed experimentally, opening new perspectives. For rapid flows on flat and smooth boundaries, our study raises several important issues. First, our measurements reveal that the contribution of the side-wall friction force with respect to the basal friction one increases with increasing slope angle and can grow up to about 25% of the latter. Direct measurements of the side-wall friction with force sensors would be useful. Second, the results reported here concerns rapid flows on smooth incline. We may legitimately wonder whether rapid flows over a rough basal substrate would exhibit similar features. Third, experiments with longer chute allowing to produce SFD flows at large angle would be definitively of great interest to consolidate our findings.

Finally, our work may have implications for geophysical granular flows. Our experimental measurements indicate that the basal friction is weakened when the particle hold-up is increased. This result may provide some clues to understand why some massive rock avalanches or pyroclastic flows exhibit unexpected long run-out distances [30].

**ACKNOWLEDGEMENTS**

We thank the two anonymous referees for stimulating reviews. We thank Jean-Louis Fruquière and Cyrille Guillot for their technical assistance. This research was financed by the French National Research Institute for Sustainable Development - IRD (O.R.), the French Government Laboratory of Excellence initiative ANR-10-LABX-0006 (O.R., S.V.D.W., L.C.), the French Research National Agency project ANR-16-CE01-0005 (A.V., R.D.), and the ERC contract ERC-CG-2013-PE10-617472 SLIDEQUAKES (A.M.). This is Laboratory of Excellence ClerVolc contribution XXX.

[1] GDR MiDi, On dense granular flows. The European Physical Journal E **14**, 341-365 (2004). Doi: 310.1140/epje/i2004-10025-10021.

[2] R. Delannay, A. Valance, A. Mangeney, O. Roche and P. Richard, Granular and particle-laden flows: from laboratory experiments to field observations. Journal of Physics D: Applied Physics **50**, 053001 (2017). Doi:053010.051088/051361-056463/053050/053005/053001.

[3] O. Pouliquen and Y. Forterre, Friction law for dense granular flows: application to the motion of a mass down a rough inclined plane. Journal of Fluid Mechanics **453**, 133–151 (2002).

[4] Y. Forterre and O. Pouliquen, Flows of Dense Granular Media. Annual Review of Fluid Mechanics **40**, 1–24 (2008).

[5] P.-Y. Lagrée, L. Staron, and S. Popinet, The granular column collapse as a continuum: validity of a two-dimensional Navier–Stokes model with a µ(I)-rheology. Journal of Fluid Mechanics **686**, 378–408 (2011).

[6] J.M.N.T. Gray and A.N. Edwards, A depth-averaged µ(I)-rheology for shallow granular free-surface flows. Journal of Fluid Mechanics **755**, 503–534 (2014).

[7] J.T. Jenkins, Boundary conditions for rapid granular flows: Flat, frictional walls. Journal of Applied Mechanics **59**, 120 (1992).

[8] M.Y. Louge, Computer simulations of rapid granular flows of spheres interacting with a flat, frictional boundary. Physics of Fluids **6**, 2253 (1994).

[9] M. Y. Louge and S. C. Keast, On dense granular flows down flat frictional inclines. Physics of Fluids **13**, 5 (2001). Doi: 10.1063/1061.1358870.

[10] A. J. Holyoake and J. N. McElwaine, High speed granular chute flows. Journal of Fluid Mechanics 710, 35-71 (2012). Doi:10.1017/jfm.2012.1331.

[11] J. Heyman, P. Boltenhagen, R. Delannay and A. Valance, Experimental investigation of high speed granular flows down inclines. EPJ Web of Conferences **140**, 03057 (2017). Doi: 03010.01051/epjconf/201714003057.

[12] Y. Forterre and O. Pouliquen, Longitudinal vortices in granular flows. Physical Review Letters **86**, 5886-5889 (2001).

[13] T. Börzsönyi and R.E. Ecke, Rapid granular flows on a rough incline: Phase diagram, gas transition, and effects of air drag. Physical Review E **74**, 061301 (2006). Doi: 061310.061103/PhysRevE.061374.061301.

[14] T. Börzsönyi, R.E. Ecke and J.N. McElwaine, Patterns in Flowing Sand: Understanding the Physics of Granular Flow. Physical Review Letters **103**, 178302 (2009). Doi: 178310.171103/PhysRevLett.178103.178302.

[15] N. Brodu, R. Delannay, A. Valance and P. Richard, New patterns in high-speed granular flows. Journal of Fluid Mechanics **769**, 218-228 (2015). Doi:210.1017/jfm.2015.1109.

[16] N. Brodu, P. Richard and R. Delannay, Shallow granular flows down flat frictional channels: Steady flows and longitudinal vortices. Physical Review E **87**, 022202 (2013). Doi: 022210.021103/PhysRevE.022287.022202.

[17] N. Taberlet, P. Richard, J.T. Jenkins and R. Delannay, Density inversion in rapid granular flows: the supported regime. European Physical Journal E **22**, 17-24 (2007). Doi: 10.1140/epje/e2007-00010-00015.

[18] L.E. Silbert, D. Ertas, G.S. Grest, T.C. Halsey, D. Levine and S.J. Plimpton, Granular flow down an inclined plane: Bagnold scaling and rheology. Physical Review E **64**, 051302 (2001). Doi: 051310.051103/PhysRevE.051364.051302.

[19] T. Faug, P. Childs, E. Wyburn and I. Einav. Standing jumps in shallow granular flows down smooth inclines. Physics of Fluids **27**, 073304 (2015). Doi: 10.1063/1.4927447.

[20] Y. Zhu, A. Valance and R. Delannay, High-speed confined granular flows down smooth inclines: scaling and wall friction laws. Granular Matter **22,** 82 (2020). Doi :10.1007/s10035-020-01053-7.

[21] S. Nasuno, A. Kudrolli, A. Bak and J.P. Gollub, Time-resolved studies of stick-slip friction in sheared granular layers. Physical Review E **58**, 2161–2171 (1998).

[22] J. Åström, H. Herrmann and J. Timonen, Granular packings and fault zones. Physical Review Letters **84**, 638–641 (2000).

[23] W. Losert, J.-C. Géminard, S. Nasuno, and J.P. Gollub, Mechanisms for slow strengthening in granular materials. Physical Review E **61**, 4060–4068 (2000).

[24] H. Lastakowski, J.-C. Géminard and V. Vidal, Granular friction: triggering large events with small vibrations. Scientific Reports **5**, 13455EP (2015).

[25] O. Hungr and N.R. Morgenstern, Experiments on the flow behaviour of granular materials at high velocity in an open channel. Géotechnique **34**, 405-413 (1984).

[26] G. Saingier, S. Deboeuf and P.-Y.Lagrée. On the front shape of an inertial granular flow down a rough incline. Physics of Fluids **28**, 053302 (2016). Doi: doi.org/10.1063/1.4948401.

[27] P. Jop, Y. Forterre and O. Pouliquen, Crucial role of side walls for granular surface flows: consequences for the rheology. Journal of Fluid Mechanics **541**, 167-192 (2005).

[28] N. Taberlet, P. Richard, A. Valance, W. Losert, J.M. Pasini, J.T. Jenkins and R. Delannay, Superstable granular heap in a thin channel. Phys. Rev. Lett. 91 (26), 264301 (2003).

[29] P. Jop, Y. Forterre and O. Pouliquen, A constitutive law for dense granular flows. Nature **441**, 727-730 (2006).

[30] A. Lucas, A. Mangeney and J.-P. Ampuero, Frictional velocity-weakening in landslides on Earth and on other planetary bodies. Nature Communications **5**, 3417 (2014). Doi: 3410.1038/ncomms4417.